\documentclass[11pt,twoside]{article}
\usepackage[activeacute,spanish]{babel}
\usepackage{waaa08-regular-uk}

\usepackage{graphicx}
\usepackage[T1]{fontenc} 
\usepackage{latexsym}
\usepackage{verbatim}

\begin{document}
\def\tablename{Tabla}%

\vskip 1.0cm
\markboth{Alejandro Gangui \& Eduardo L. Ortiz}%
{First echoes of relativity in Argentine astronomy}

\pagestyle{myheadings}

\vskip 0.3cm
\title{First echoes of relativity in Argentine astronomy} 

\author{Alejandro Gangui$^{1,2}$, Eduardo L. Ortiz$^{3}$}

\affil{%
  (1) IAFE -- Instituto de Astronom{\'\i}a y F{\'\i}sica del Espacio -- CONICET, \\
      Ciudad Universitaria, Buenos Aires, Argentina \\
  (2) CEFIEC -- Facultad de Ciencias Exactas y Naturales -- UBA, \\
      Ciudad Universitaria, Buenos Aires, Argentina\\
  (3) Imperial College London, \\ South Kensington Campus, London SW7 2AZ, England\\
}


\begin{abstract}

In this note we consider the attitude of astronomers in Argentina in connection with the new problems posed by
relativity theory, before and after General Relativity was presented in its final form. We begin considering, very
briefly, the sequence of ``technical'' publications related to relativity that appeared in Argentina and use it to
attempt to identify who were the relativity leaders and authors in the Argentina scientific community of the
1910-1920s. Among them there are natives of Argentina, permanent resident scientists, and occasional foreign
visitors. They are either academic scientists, or high school teachers; we leave aside the {\it philosophers} and
the {\it aficionados}. For the main characters we discuss, very briefly again, the scientific facts and publications
they handled, the modernity of their information and the ``language'' they use to transmit their ideas to their
readers.

Finally, we consider astronomers proper; first Charles Dillon Perrine, an astronomer interested in astrophysics,
contracted by the government of Argentina in the USA as director of its main observatory. He became interested in
testing the possible deflection of light rays by the Sun towards 1912; his Argentine expedition was the first to
attempt that test. Perhaps Perrine was not so much interested in Einstein's formulation of relativity theory,
which then was perceived as very far away from his own field of research, as in testing the particular astronomical
effects it predicted. In any case, he attempted to do so with the acquiescence and financial support of the
Argentine state, and as a leading member of its official scientific elite. We briefly contrast his very specific and
strictly scientific efforts with those of our second astronomer, Jos\'e Ubach, SJ, a secondary school teacher of
science at a leading Buenos Aires Catholic school who reported in response to Eddington's expedition. Finally,
our third astronomer is F\'elix Aguilar, a leading scientist with a more definite interest in astrometry, who
made an effort to contribute to the public understanding of Einstein's theories in Argentina in 1924, when
Einstein's visit to Argentina had become a certainty.

\end{abstract}

\section{First publications}

Even some years after 1905, it was only a few authors that discussed advanced dynamics, electron theory and
radiation in Argentina. In their works they did not necessarily refer to the 1905 work of Einstein. We follow (Ortiz
1995) to present a list of ``technical'' papers connected with relativity. The main responses are those of Lepiney
(1906-8), who makes reference to work of Max Abraham on electron dynamics, and again (Lepiney 1907), now with a
general discussion of dynamics with a velocity-dependent mass. Also Broggi (1909) discussed Lorentz's
electrodynamics and included a mathematical analysis of the works of Minkowski. Following the early experiments of
J. J. Thomson, physicists knew that the motion of an electron was modified in the presence of an electromagnetic
field, which could be interpreted as an increment of its mass. Lorentz's aether theory and descriptions of the
behaviour of the electron, compatible with Maxwell's equations, were also discussed at the time. All these ideas, as
well as Abraham's description of the electron as a perfect sphere with surface charge, were descriptions that agreed
with ordinary `common sense'.

In 1910 Vito Volterra delivered a lecture at the Sociedad Cient{\'\i}fica Argentina (the Argentine Scientific
Society; SCA herein) in Buenos Aires; in (Volterra 1910) he discussed the now well-known paper which started the
relativity revolution and, with it, Einstein's {\it annus mirabilis} (see Gangui 2007). In the decade of 1910 Jakob
Johann Laub, a physicist of Polish origin, trained in Germany and hired by the Physics Institute at La Plata,
offered lectures and gave courses connected with the theory of relativity. Pyenson (1985), who has studied Laub's
personality in detail, has indicated that he was the first physicist to co-author a paper with Einstein, and also
suggested that a set of lectures on relativity theory given by him at La Plata may have been the first course on
that subject given in the Americas. Once in Argentina Laub translated into Spanish some results obtained in
Europe. In (Laub 1912) he discussed briefly optical effects in moving bodies in a paper published in the {\it
  Anales} of the SCA. The French physicist Camilo Meyer, a former secondary school companion of Henri Poincar\'e in
France, delivered a series of optional courses on mathematical physics at the University of Buenos Aires in 1910-15;
even if he did not specialize in relativity, his courses mentioned recent research in physics, including Kaufmann's
experiments on the velocity dependence of the electron mass; he also made reference to Einstein's work without
entering into details. Again, between 1916 and 1919, Laub (1916; 1919) considered physical and philosophical
questions connected with relativity theory from the point of view of a physicist. In the same years there were also
translations or adaptations of general articles that reflected an interest on relativity theory; mostly, they were
taken from the foreign press or from popular science journals.

\section{Perrine's involvement in the first attempts to verify observationally Einstein's ideas}

In the meantime, Einstein laboriously worked to complete his theory and, eventually, incorporate gravitation to his
new relativistic framework. In (Einstein 1907) he made his first statement of what later became known as the
equivalence principle. In it he assumes ``the complete physical equivalence of a gravitational field and a
corresponding acceleration of the reference system.'' Einstein then combined this principle with key assumptions of
Special Relativity and was able to predict that clocks would run at slightly different rates if located in different
places within an inhomogeneous gravitational field (smaller rates for strongest fields). Another conclusion he
derived, which turned out to be a most important one for the acceptance of the theory, was that light-rays would
bend in a gravitational field. Einstein developed his thoughts in an article published in 1911, in which he was
looking for a new framework that would allow him not to postulate, but to derive the equivalence principle, and led
to a more general relativity principle as compared to his 1905 proposal. It is in this work that Einstein combined
his equivalence principle with Newton's gravitational theory and computed, wrongly, the deviation suffered by a
light-ray of a far-away background star, as it travelled close to the Sun's limb, towards an observer on Earth. He
gave the value of 0.87'' for the bending of light in the gravitational field of the Sun, which he would later
revise. For both, the gravitational red-shift and the bending of light-rays, Einstein found a useful collaborator in
Erwin Freundlich, a young astronomer who became interested in putting these new ideas to test by astronomical means.

Perrine, of Lick Observatory, California, was a world class astronomer with a solid reputation for his achievements
in his field (see Hodge 1977); in 1909 he accepted the position of director of the Argentine National Observatory,
C\'ordoba (see Landi Dessy 1970; Bernaola 2001). In (Perrine 1923; 1931) he has described with concision, but
accurate details, his early involvement with the testing of relativity. Let us recount the main points. In a brief
visit of Perrine to Berlin, in 1911, young Freundlich asked him for advice on Einstein's deflection problem; the
matter involved an eclipse observation, which was an area in which Perrine was a world leader. The topic was also
close to Perrine's past interests and, as a consequence of Freundlich requests, he made early efforts to test
relativity in several eclipse expeditions he conducted from Argentina. Perrine's attempts began with the Brazil
total solar eclipse of 1912, which he observed as head of the Argentine mission; sadly, as it often happens with
eclipses, adverse meteorological conditions prevented him from making good observations and producing the required
results. Laub also travelled to Brazil for the observation of the eclipse, but his interests were not directly
connected with relativity, but with atmospheric electricity. The same happened to Perrine on a second Argentine
expedition organized a couple of years later, this time to Russia. Perrine's old friend and colleague at Lick,
William Wallace Campbell, who had also became interested in the testing, was also in Russia, as well as
Freundlich. The latter, a German, was prevented from making observations on account of the First World War. Bad
weather again, made it impossible for anyone to produce accurate results. As it is well known, in 1919 it was Arthur
S. Eddington who resolved the matter.

\section{``Post-eclipse'' publications}

After the November 1919 announcement of Eddington's eclipse results, notes on Einstein's ideas attracted
considerable public interest and articles appeared in journals of different levels all over the world. This was also
the case in Argentina, where a number of lectures and articles, neither fully technical nor entirely at popular
level, appeared. Some of them clearly stated that they were not expositions for the expert, or scientific
innovations, but contributions to satisfy the interests of the general reader, as for example the astronomer Aguilar
(1924) made quite clear.

In addition to the ones cited above from a much larger list, the main authors involved in disseminating the new
ideas of relativity in the Argentine community included visitors such as Blas Cabrera, Richard Gans (director of La
Plata's Institute of Physics), or Georg Friedrich Nicolai (visiting professor of physiology in the University of
C\'ordoba); stable members of Argentina's academic or education circles such as Aguilar, engineers Enrique Butty and
Jorge Duclout, physicists Jos\'e Collo and Te\'ofilo Isnardi, writer and poet Leopoldo Lugones, mathematician Julio
Rey Pastor and the astronomer and Jesuit priest and teacher Jos\'e Ubach. A number of philosophical and
pseudo-philosophical interpretations of relativity found also a fertile soil in the Argentina of the 1920's (see
As\'ua \& Hurtado de Mendoza 2006). More details can be found in (Ortiz 1995; Gangui \& Ortiz 2005; and Ortiz \&
Rubinstein 2008).

In Argentina relativistic ideas were propagated through journals associated with scientific societies, university,
professional associations or student union's, as well as by literary journals. Among others: {\it Anales de la
  Sociedad Cient{\'\i}fica Argentina}, {\it Anales de la Universidad de Buenos Aires}, {\it Revista Humanidades
  (University of La Plata)}, {\it Revista T\'ecnica}, {\it Bolet{\'\i}n del Centro Naval}, {\it Revista
  Polit\'ecnica} (later {\it Revista del Centro de Estudiantes de Ingenier{\'\i}a}, or {\it CEI}), {\it Verbum}
(journal of the Buenos Aires Humanities Student's Union, the Centro de Estudiantes de Filosof{\'\i}a y Letras),
{\it Revista de Filosof{\'\i}a} and {\it Nosotros}.

However, an interesting and rather unusual channel for the diffusion on {\it Einsteiniana} in Argentina was {\it La
  Vida Literaria}, a fringe literary journal of limited circulation, produced by left-wing writers and poets, which
was responsible for the publication of what has been called Einstein's {\it in\'edito}: the philosophically oriented
text of the lecture Einstein intended to use to open his courses at the University of Buenos Aires, but which
somebody persuaded him to leave aside ``to keep everybody happy'' (see Gangui \& Ortiz 2008).

\section{Astronomy: the Collo-Isnardi-Aguilar paper and the testing of relativity}

Some of the references mentioned above touched upon certain topics of astronomy, but did not consider them in any
detail. The first thorough description of the astronomical testing of Einstein's ideas in Argentina, as (Ortiz 1995)
has shown, is a neglected contribution of Father Jos\'e Ubach. A science teacher at Colegio del Salvador, Buenos
Aires, Ubach had received training in Catalu\~na. He reviewed the results of Eddington's expedition in (Ubach 1920)
immediately after the former published his results. His views were critical and circumspect, but on the whole
balanced. His main point being that the results of the 1919 observations were important but, on account of the
complexity of the observations, not yet definitive. Within the Argentine scientific community, Ubach's views
reflected a more open attitude of the Catholic Church in Europe vis-\`a-vis contemporary scientific research, and a
further manifestation of the movement who supported becoming more directly involved in it.

Some four years later, in 1924, in preparation for Einstein's arrival in Argentina, F\'elix Aguilar published a note
on the results of the same expedition in {\it Bolet{\'\i}n del Centro Naval}, the journal of the navy officers club
(Centro Naval).  His review is the third in a set of three articles on relativity theory; the first two were written
by Jos\'e Collo and by Te\'ofilo Isnardi, respectively. These three authors were among the young most promising
Argentine researchers of the time. As we pointed out before, the articles were neither technical nor popular,
addressed to ``those who, without being experts, possess enough knowledge to become interested in some of the
details of the development of the theory'' (Collo, Isnardi \& Aguilar, 1923-24). Their reviews, they said, were
motivated by cultural considerations; that is, strictly, they were not ``scientific'' papers.

We will only highlight some of the main ingredients of the first two of these three papers, and then concentrate a
bit more on the third one, the astronomical review by Aguilar. Collo was in charge of the first part, dealing with
``preliminaries'' on the special theory, from Galilean mechanics up to Einstein's conceptions of time, simultaneity,
the postulates of Special Relativity, and ending with Lorentz transformations and Minkowski geometric
representations (Collo 1923). In the second paper Isnardi focused on General Relativity and gave a discussion up to
the theory's predictions regarding the deflection of light in a homogeneous gravitational field and also the
resulting gravitational red-shift of light propagating in an inhomogeneous field as that of the Sun. In the second
half of his contribution, he computed geodesics in Schwarzschild spacetime getting the classical value of 43'' per
century for the anomalous perihelion advance of the planet Mercury and the 1.74'' deflection of light-rays of
background stars passing close to the Sun (Isnardi 1923).

The third paper (Aguilar 1924) began with a historical review of the question of the anomalous perihelion advance of
Mercury (to which Perrine, with his observations and search of a possible intra-mercurial planet and celestial
photography, had contributed substantially), and Einstein's interpretation of this phenomenon. Aguilar then
discussed the observations related to the verification of the second classical test of General Relativity: the
deflection suffered by background starlight passing close to the limb of the Sun. In this part, he reviewed briefly
the 1914 eclipse, but did not mention Perrine's work or the Argentine expeditions of 1912 and 1914. He gave a
detailed account of the famous British eclipse expeditions of 1919 to Sobral, in Northern Brazil, and to the island
of Pr{\'\i}ncipe, near Africa, which confirmed Einstein's predictions. For these, as well as for the following Lick
Observatory eclipse expedition in Australia of September 21st, 1922, he included tables and diagrams of the shifts
in the position of many background stars, quoted even with error bounds, and photographs of the eclipsed Sun and of
the experimental setting. Aguilar's article finished with three pages in which he explained the extraordinary
difficulties involved in trying to test the third classical prediction of General Relativity, namely, the tiny
red-shift of the Sun light spectrum due to the gravitational field of our star. He quoted the analysis Freundlich
and others performed on Sun plates obtained for previous studies of the Sun; however, he failed to remark that these
plates had come to Freundlich's hands through the generous intervention of Perrine, the astronomer from local
C\'ordoba. Aguilar carefully emphasised the difficult problem of differentiating the Doppler shifts, kinematical in
origin, from the gravitational shifts. As Ubach before him, Aguilar concluded that the situation was not clear,
neither in favour nor against General Relativity predictions, and that Einstein's theory was pushing experimental
observations to their technical limit.

\section{Final remarks}

New developments in mathematics and in theoretical physics attracted attention in Argentina from at least the last
third of the nineteenth century; from the early part of the twentieth century there was an interest in the new
theory of ``quanta'', and later in Einstein's relativity theory. Argentina's economical prosperity made it possible
to attract to its universities and advanced institutions scientists with a remarkable record. One of them, Charles
Dillon Perrine, director of the C\'ordoba National Observatory, played an interesting role in the earlier efforts to
verify Einstein's theory, personally and through his advice to others, Freundlich, among them. The importance of
these attempts, understandably, may not have been as clear then as they were after 1919. However, even as late as
1926, after Einstein's visit to Argentina in 1925 and after the publication of (Perrine 1923), such perception is
still absent in both (Aguilar 1924) and in the official SCA's history of astronomy in Argentina for the period
1872-1922 (Chaudet 1926). In his review, Chaudet makes reference to the 1912 and 1914 eclipse expeditions of the
C\'ordoba Observatory, of which he was an employee, but without any reference to Perrine's attempts in connection
with relativity theory (Chaudet 1926, p. 72). In any case, Perrine attempted to prove or disprove relativity with
the acquiescence and financial support of the Argentine state, and as a leading member of its official scientific
elite. Ubach's interesting paper had also gone into oblivion. Clearly, there was some lack of communication at the
highest scientific levels of the astronomical community in the Argentina of the mid 1920s.

\agradecimientos 
The work of A.G. was partially supported by grants PIP-6332 (CONICET) and UBACyT X439 (University of Buenos Aires).

\begin{referencias}

\reference Aguilar, F. 1924, `Teor{\'\i}a de la Relatividad', {\it Bolet{\'\i}n del Centro Naval} 41 (445): 747-762.

\reference As\'ua, M. de, \& Hurtado de Mendoza, D. 2006, {\it Im\'agenes de Einstein: relatividad y cultura en el
  mundo y en la Argentina}, Buenos Aires: Eudeba.

\reference Bernaola, O. 2001, {\it Enrique Gaviola y el Observatorio Astron\'omico de C\'ordoba. Su impacto en el
  desarrollo de la ciencia argentina}. Buenos Aires: Saber y Tiempo.

\reference Broggi, U. 1909, `Sobre el principio electrodin\'amico de relatividad y sobre la idea de tiempo', {\it
  Revista Polit\'ecnica} 10 (86): 41-44.

\reference Chaudet, E. 1926, {\it La Evoluci\'on de la Astronom{\'\i}a durante los \'ultimos cincuenta a\~nos
  (1872-1922)}, Buenos Aires: SCA.

\reference Collo, J. 1923, `Teor{\'\i}a de la Relatividad', {\it Bolet{\'\i}n del Centro Naval} 41 (442): 264-284.

\reference Collo, J., Isnardi, T. \& Aguilar, F. 1923, `Teor{\'\i}a de la Relatividad', {\it Bolet{\'\i}n del Centro
  Naval} 41 (442): 263.

\reference Einstein, A. 1907, `\"Uber das Relativit\"atsprinzip und die aus demselben gezogene Folgerungen', {\it
  Jahrbuch der Radioaktivitaet und Elektronik} 4; translated `On the relativity principle and the conclusions drawn
from it', in {\it The collected papers of Albert Einstein}. Vol. 2: The Swiss years: writings, 1900--1909 (Princeton
University Press, Princeton, NJ, 1989), Anna Beck translator.

\reference Gangui, A. (ed.) 2007, {\it El universo de Einstein: 1905 -- annus mirabilis -- 2005}, Buenos Aires:
Eudeba. [http://arxiv.org/abs/0705.4266]

\reference Gangui, A., \& Ortiz, E. L. 2005, `Albert Einstein visita la Argentina (Marzo-Abril 1925: cr\'onica de un
mes agitado)', {\it Todo es Historia} 454, pp. 22-30, May issue. [http://arxiv.org/abs/physics/0506052]

\reference Gangui, A., \& Ortiz, E. L. 2008, `Einstein's unpublished opening lecture for his course on Relativity
Theory in Argentina, 1925', {\it Science in Context} 21 (3): 435-450.

\reference Hodge, J. 1977, `Charles Dillon Perrine and the transformation of the Argentina National Observatory',
           {\it Journal for the History of Astronomy} 8: 12-25.

\reference Isnardi, T. 1923, `Teor{\'\i}a de la Relatividad', {\it Bolet{\'\i}n del Centro Naval} 41 (443): 413-449.

\reference Landi Dessy, J. 1970, `Charles Dillon Perrine y el desarrollo de la astrof{\'\i}sica en la Rep\'ublica
Argentina', {\it Bolet{\'\i}n de la Academia Nacional de Ciencias} 48: 219-234.

\reference Laub, J. J. 1912, `Noticia sobre `Los efectos \'opticos en medios en movimiento'', {\it Anales de la
  Sociedad Cient{\'\i}fica Argentina} 73: 38-46.

\reference Laub, J. J. 1916, `Los teoremas energ\'eticos y los l{\'\i}mites de su validez', {\it Revista de
  Filosof{\'\i}a} 4: 59-73.
 
\reference Laub, J. J. 1919, `Qu\'e son espacio y tiempo?', {\it Revista de Filosof{\'\i}a} 9: 386-405.

\reference Lepiney, P. de (1906-8), `Los electrones y las radiaciones', {\it Revista T\'ecnica} (Buenos Aires) XII
(1906-1907): 109-114; Ib{\'\i}d. 144-149; XIII (1907-1908): 169-171.

\reference Lepiney, P. de 1907, `La din\'amica sin el segundo principio de Newton', {\it Anales de la Universidad de
  Buenos Aires} 7: 56-57.

\reference Ortiz, E. L. 1995, `A convergence of interests: Einstein's visit to Argentina in 1925',
{\it Ibero-Americanisches Archiv} 20: 67-126.

\reference Ortiz, E. L. \& Rubinstein, H. 2008, `La F{\'\i}sica en la Argentina entre 1900 y 1966: algunos
condicionantes exteriores a su desarrollo', {\it Saber y Tiempo} (in press).

\reference Perrine, C. D. 1923, `Contribution to the history of attempts to test the theory of relativity by means
of astronomical observations', {\it Astronomische Nachrichten} 219: 281-4.

\reference Perrine, C. D. 1931, `Fundaci\'on del Observatorio Nacional Argentino y sus objetivos', {\it Anales de la
  Sociedad Cient{\'\i}fica Argentina} 111: 281-294.

\reference Pyenson, L. 1985, {\it Cultural Imperialism and Exact Sciences: German Expansion Overseas 1900-1930}, New
York: Peter Lang, NY.

\reference Ubach, S.J., J. 1920, {\it La teor{\'\i}a de la relatividad en la f{\'\i}sica moderna: Lorentz,
  Minkowski, Einstein}, Buenos Aires: Amorrortu.

\reference Volterra, V. 1910, `Espacio, tiempo i masa', {\it Anales de la Sociedad Cient{\'\i}fica Argentina} 70:
223-243.

\end{referencias}
\end{document}